\documentclass{article}

\usepackage{amsmath}
\usepackage[dvips]{geometry}
\usepackage{fullpage}
\usepackage[T1]{fontenc}
\usepackage{graphicx}
\usepackage{SIunits}

\newcommand{\J}{\mathrm{j}}  
\newcommand{\D}{\mathrm{d}}  
\newcommand{\E}{\mathrm{e}}  

\newcommand{\mt}{M_{\text{t}}}

\begin{document}

\bibliographystyle{unsrt}

\title{Information in spinning sound fields\footnote{This paper is
    based on work which was to have been presented at the 159th
    Meeting of the Acoustical Society of America, Baltimore, 2010.}}
\author{Michael Carley} 

\maketitle

\begin{abstract}
  The information content of a spinning sound field is analyzed using
  a combination of exact and asymptotic results, in order to set
  limits on how accurately source identification can be carried
  out. Using a transformation of the circular source to an exactly
  equivalent set of line source modes, given by Chebyshev polynomials,
  it is found that the line source modes of order greater than the
  source wavenumber generate exponentially small fields. Asymptotic
  analysis shows that the remaining, lower order, modes radiate
  efficiently only into a region around the source plane, with this
  region shrinking as the mode order is increased. The results explain
  the ill-conditioning of source identification methods; the
  successful use of low order models in active noise control; and the
  low radiation efficiency of subsonic jets.  
\end{abstract}

\section{Introduction}
\label{sec:intro}

Source identification, the problem of determining an acoustic source
from field measurements, has been attempted using a number of
approaches in various different technologies. This paper examines the
problem of identifying the source which generates a spinning acoustic
field. The source which generates such fields can be represented as a
set of modes which vary with azimuth on a circular disk, whether or
not the source includes a spinning element.  Examples include rotating
systems such as cooling
fans~\cite{gerard-berry-masson05a,gerard-berry-masson05b}, helicopter
rotors~\cite{brooks-marcolini-pope87,marcolini-brooks92} duct
terminations such as aircraft engine
intakes~\cite{holste-neise97,farassat-nark-thomas01,lewy05,lewy08,%
  castres-joseph07a,castres-joseph07b} and
jets~\cite{jordan-schlegel-stalnov-noack-tinney07} if a jet is
modelled as a distribution of disk-shaped sources.

There are two broad categories of problem where source identification
is required, corresponding to `forward' and `backward' projection of
the field. In the forward problem, the aim is to estimate the source
distribution accurately enough to allow the field to be predicted at
positions other than the original measurement points. This has been
done, for example, in extracting source parameters from near-field
measurements of propeller noise, with the parameters then being used
to calculate far-field noise\cite{peake-boyd93}. 

In backward projection the problem is determination of the source
proper from field measurements. This might be done in order to decide
on noise control measures or because the acoustic source corresponds
to some other physical variable of interest. In the first case, the
aim is usually to find the acoustic source strength distributed over
the source region, such as a rotor
disk\cite{gerard-berry-masson05a,gerard-berry-masson05b} or the
termination of a
duct\cite{lewy05,lewy08,castres-joseph07a,castres-joseph07b}. An
example of the second application is the study of noise generation by
turbulent jets\cite{jordan-schlegel-stalnov-noack-tinney07}, where the
aim is to determine the fluid-dynamical mechanisms which give rise to
the acoustic source. 

In any case, it is well known that the problem of source
identification is poorly-conditioned, meaning that small measurement
errors can give rise to very large changes in the estimated
source. Previous studies of the structure of spinning
fields\cite{chapman93,carley99,carley00,carley01} have shown that the
field decays exponentially away from the source. This means that in
the forward projection problem, errors in the estimated acoustic
source will decay and the predicted acoustic field may well be quite
accurate, even if the source is not well recovered. On the other hand,
in the backward projection problem, the exponential decay moving away
from the source corresponds to exponential growth moving towards it,
leading to large errors in the estimated source. 

In this paper, the radiated field from a disk source is analyzed to
examine how much information about the source can actually be detected
in the field.In a previous paper on a possible method for source
identification\cite{carley09}, it was shown that the far-field noise
is band-limited Fourier transform of a line source which is exactly
equivalent to the disk source. In this paper, without recourse to far
field approximations, it is possible to establish fundamental limits
on the number of degrees of freedom of a field, limits which are
determined by the source frequency. The implications of the analysis
are discussed with respect to some real source identification and
radiation prediction problems.

\section{Analysis}
\label{sec:analysis}

\begin{figure}
  \centering
  \includegraphics{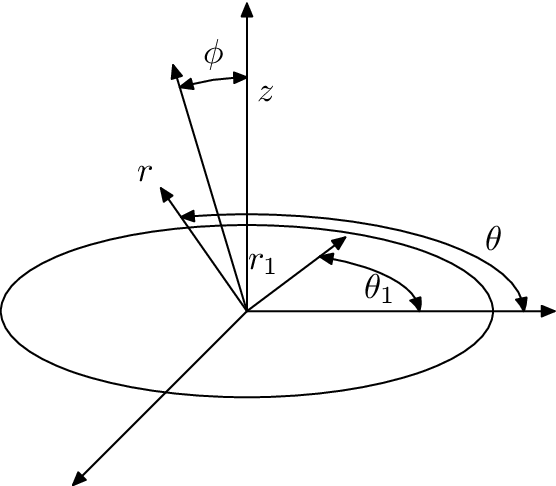}
  \caption{Coordinate system for radiation prediction}
  \label{fig:disk} 
\end{figure}

The problem considered is that of the field generated by an
azimuthally varying distribution of monopoles with strength
$s(r_{1},\theta_{1})$ given by the Rayleigh
integral\cite{goldstein74,chapman93}
\begin{align}
  \label{equ:full}
  p(r,\theta,z,\omega) &= \int_{0}^{1}\int_{0}^{2\pi}
  s(r_{1},\theta_{1}) \frac{\E^{\J k R}}{4\pi
    R}\,\D\theta_{1}\,r_{1}\,\D r_{1},\\
  R^{2} &= r^{2} + r_{1}^{2} - 2r r_{1}\cos(\theta-\theta_{1}) +
  z^{2}, \nonumber
\end{align}
where the source is distributed over the unit disk in the plane $z=0$,
variables of integration have subscript $1$ and the coordinate system
is shown in Figure~\ref{fig:disk}. The wavenumber $k=\omega/c$, and
$c$ is the speed of sound.

Taking one azimuthal mode of the source distribution,
$s(r_{1},\theta_{1})=s_{n}(r_{1})\exp\J n\theta_{1}$, the radiated
field for one mode can be written $p=p_{n}\exp\J n\theta$:
\begin{align}
  \label{equ:mode}
  p_{n}(r,z) &= 
  \int_{0}^{1}\int_{0}^{2\pi} s_{n}(r_{1})
  \frac{\E^{\J (k R-n\theta_{1})}}{4\pi R}\,\D\theta_{1}\,r_{1}\,\D
  r_{1},\\
  R^{2} &= r^{2} + r_{1}^{2} - 2r r_{1}\cos\theta_{1} + z^{2}.\nonumber
\end{align}
The integral of Equation~\ref{equ:mode} has been extensively studied
due to its relevance to rotor acoustics and, under suitable
conditions, as a good approximation to radiation from ducts. Many
problems in source
identification\cite{gerard-berry-masson05a,gerard-berry-masson05b,%
holste-neise97,farassat-nark-thomas01,lewy05,lewy08,%
castres-joseph07a,castres-joseph07b} can be viewed as attempts to
recover the source term $s_{n}(r_{1})$ from measurements of $p_{n}$.

The remainder of this paper consists of an analysis of the integral of
Equation~\ref{equ:mode}, which will establish limits on the
information about the source which is radiated into the acoustic
field, thereby fixing how accurately a source can be identified. The
results are also applicable to the question of the detail with which a
source need be specified in order to accurately predict the acoustic
field, and to that of the radiation efficiency of jets.


\subsection{Equivalent line source}
\label{sec:line}

\begin{figure}
  \centering
  \includegraphics{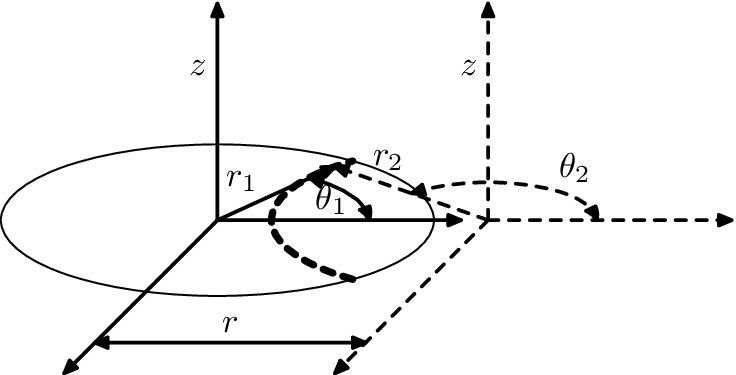}
  \caption{Coordinates for transformation to equivalent line source}
  \label{fig:coordinates}
\end{figure}

The first step in the analysis is to transform the disk source into a
line source which generates (exactly) the same acoustic field. This is
a transformation which has been used, in the axisymmetric case, in
studies of transient radiation from
pistons\cite{oberhettinger61a,pierce89}, and, with azimuthal
variation, in studies of rotor
acoustics\cite{chapman93,carley99,carley00,carley01}. The first stage
is to switch from the source-centred cylindrical coordinates
$(r,\theta,z)$ of Figure~\ref{fig:disk} to the observer-centred
coordinates $(r_{2},\theta_{2},z)$ of
Figure~\ref{fig:coordinates}. Under this transformation,
Equation~\ref{equ:mode} becomes, for $r>1$:
\begin{align}
  \label{equ:transformed}
  p_{n}(k,r,z) &= \int_{r-1}^{r+1} \frac{\E^{\J kR}}{R}
  K(r,r_{2})r_{2}\,\D
  r_{2},\\
  R &= \left(r_{2}^{2} + z^{2}\right)^{1/2},\nonumber\\
  \label{equ:kfunc}
  K(r,r_{2}) &= \frac{1}{4\pi}
  \int_{\theta_{2}^{(0)}}^{2\pi-\theta_{2}^{(0)}} \E^{-\J
    n\theta_{1}}s_{n}(r_{1})\,\D\theta_{2},
\end{align}
where the source function $K(r,r_{2})$ depends on $r$, the observer
lateral separation, but is independent of $z$, the axial
displacement. The coordinate systems are related by:
\begin{subequations}
  \label{equ:reverse}
  \begin{align}
    r_{1}^{2} &= r^{2} + r_{2}^{2} + 2rr_{2}\cos\theta_{2},\\
    \theta_{1} &= \tan^{-1}
    \frac{r_{2}\sin\theta_{2}}{r+r_{2}\cos\theta_{2}}.
  \end{align}
\end{subequations}
and the limits of integration in Equation~\ref{equ:kfunc} are given by
setting $r_{1}=1$:
\begin{align}
  \label{equ:theta}
  \theta_{2}^{(0)} &= \cos^{-1}\frac{1-r^{2}-r_{2}^{2}}{2rr_{2}}.
\end{align}

The function $K(r,r_{2})$ has square-root behavior at its
end-points\cite{carley09}, $r_{2}=r\pm1$, and can be expanded:
\begin{align}
  \label{equ:K:poly}
  K(r,r_{2}) &= \sum_{m=0}^{\infty}u_{m}(r)U_{m}(s)(1-s^{2})^{1/2}
\end{align}
with $s=r_{2}-r$ and $U_{m}(s)$ the Chebyshev polynomial of the
second kind.

Inserting this expansion into Equation~\ref{equ:transformed}
\begin{align}
  \label{equ:expanded}
  p_{n}(k,r,z) &= \sum_{m=0}^{\infty}u_{m}(r)I_{m}(k,r,z),\\
  \label{equ:un:basic}
  I_{m}(k,r,z) &= \int_{-1}^{1} \frac{\E^{\J k R}}{R} U_{m}(s)
  (r+s)(1-s^2)^{1/2}\, \D s,\\
  \label{equ:un:basic:xf}
  &= \int_{0}^{\pi} \frac{\E^{\J k R}}{R}(r+\cos\beta)\,
  \sin(m+1)\beta\,\,\sin\beta\,\D\beta,\\
  R^{2} &= (r+\cos\beta)^{2} + z^{2},\nonumber
\end{align}
with the transformation $s=\cos\beta$ and use of the definition of the
Chebyshev polynomial\cite{gradshteyn-ryzhik80},
$U_{m}(s)=\sin[(m+1)\beta]/\sin\beta$.

\section{Radiated field}
\label{sec:field}

The analysis of the previous section gives us a model of a spinning
acoustic field expressed in terms of an exactly equivalent line source
composed of a superposition of modes given as Chebyshev polynomials,
with the modal coefficients functions of observer radius $r$, but not
of axial displacement $z$. In this section, we use the model to draw
basic conclusions about the acoustic information which is available
for source identification.

\subsection{Cut-off modes}
\label{sec:cut:off}

The first conclusion we can draw from the integral expression for
$I_{m}$ is that modes with $m>k$ are exponentially small and can be
considered `cut-off'. For $z=0$, $I_{m}(k,r,0)$ can be evaluated
exactly using Equation~\ref{equ:un:basic:xf} and standard relations
for Bessel functions\cite{gradshteyn-ryzhik80}:
\begin{align}
  \label{equ:in:plane}
  I_{m}(k,r,0) &= \J^{m}\pi (m+1) \frac{J_{m+1}(k)}{k} \E^{\J k r}.
\end{align}
For $m+1$ large and $k<m+1$, the Bessel function $J_{m+1}(k)$ decays
exponentially, i.e. the higher modes are `cut off'. Since
$|I_{m}(r,z)|$ has its maximum in the plane $z=0$, we can further
conclude that modes with $m>k$ are cut off everywhere and cannot be
detected in the field. This is an exact result which places a first
limit on the information radiated.

\subsection{Cut-on modes}
\label{sec:sphase}

A second limit on the information available in the acoustic field can
be found by asymptotic analysis of $I_{m}$ which can be rewritten:
\begin{align*}
  I_{m} &= (Q_{m+2}(k,r,z) + Q_{-m-2}(k,r,z) \nonumber\\
  &- Q_{m}(k,r,z) - Q_{-m}(k,r,z))/4,
\end{align*}
with:
\begin{align}
  \label{equ:sp:basic}
  Q_{m}(k,r,z) &= \int_{0}^{\pi}\E^{\J k \psi(\beta)}
  \frac{r+\cos\beta}{R}\,\D\beta,\\
  \psi(\beta) &= R + \gamma\beta,\nonumber\\
  \gamma &= m/k.\nonumber
\end{align}
The integral $Q_{m}$ is in a suitable form for stationary phase
analysis\cite{bleistein-handelsman86}, which depends on finding the
stationary points of $\psi$, i.e.\ values of $\beta$ where
$\D\psi/\D\beta=0$ with $0\leq\beta\leq\pi$. Upon differentiation and
rearrangement, the condition $\D\psi/\D\beta=0$ takes the form of a
quartic equation:
\begin{align}
  \label{equ:roots}
  (\alpha^{2}-C^{2})(r+C)^{2} - \gamma^{2}z^{2} &= 0,
\end{align}
where $C=\cos\beta$ and $\alpha^{2}=1-\gamma^{2}$. To lie in the
domain of integration, the stationary phase points must be real with
$|C|<1$. This leads to the requirement that $0<\gamma<1$ and
$|z|<z_{c}$, a `cut-off' value of observer axial displacement beyond
which the phase function $\psi$ has no valid stationary points. The
two values of $C$, denoted $C_{+}$ and $C_{-}$, at the limits of $|z|$
are:
\begin{align}
  \label{equ:spoint}
  C_{\pm}(z) &= 
  \left\{
      \begin{matrix}
        &\pm\alpha, &z &= 0,\\
        &-r/4+(r^2+8\alpha^2)^{1/2}/4,\quad &|z| &= z_{c}.
      \end{matrix}
    \right.
\end{align}
Denoting $C_{c}=C_{+}(z_{c})=C_{-}(z_{c})$, the cut-off value of $z$
is:
\begin{align}
  \label{equ:zc}
  z_{c} &= (\alpha^{2}-C_{c}^{2})^{1/2}(r+C_{c})/\gamma,\\
  &\to \alpha r/\gamma,
  \quad \alpha/r\to0.\nonumber
\end{align}
Written in spherical coordinates, the asymptotic cut-off lines
$z_{c}=\alpha r/\gamma$ are rays with polar angle
$\phi=\sin^{-1}\gamma$. For completeness, we note that for $\gamma=0$,
there is no cutoff and the line source mode radiates into the whole
field with amplitude proportional to $k^{-1/2}$.

\begin{figure}
  \centering
  \includegraphics{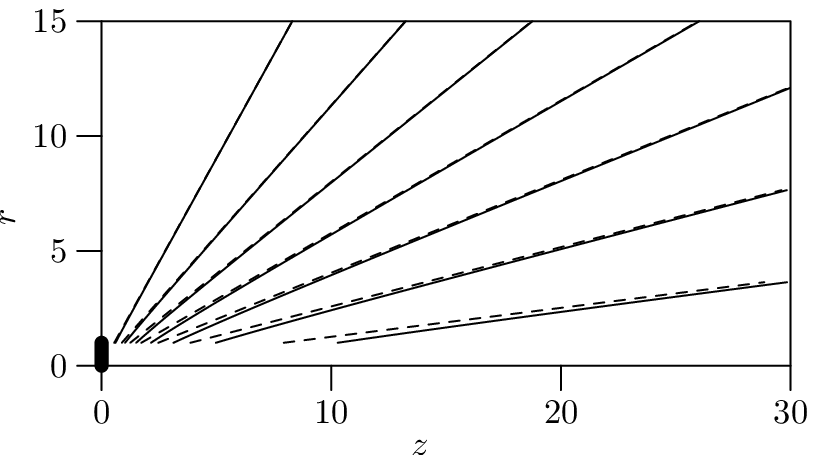}
  \caption{Cut-off lines for $\gamma=m/8$, $m=1,2,\ldots,6,7$; exact
    solution solid; $z_{c}=\alpha r/\gamma$ dashed.}
  \label{fig:cutoff}
\end{figure}

Figure~\ref{fig:cutoff} shows the cut-off lines, exact and asymptotic,
for $0<\gamma<1$. The accuracy of the asymptotic approximation for the
cut-off line is confirmed and the plot shows which radiated modes are
detectable in a given part of the acoustic field.

Using the previous results, the asymptotic behavior of the basic
integral is given by:
\begin{align}
  \label{equ:sp:ans}
  Q_{m} &\sim 
  \J^{3/2}\frac{\E^{\J k \psi_{+}}}{(kR_{+})^{1/2}}
  \left[
    \frac{2\pi}{C_{+}(r+2C_{+}) - \alpha^{2}}
  \right]^{1/2}(r+C_{+}) \nonumber\\
  &+\J^{1/2}\frac{\E^{\J k \psi_{-}}}{(kR_{-})^{1/2}}
  \left[
    \frac{2\pi}{C_{-}(r+2C_{-}) - \alpha^{2}}
  \right]^{1/2}(r+C_{-}),\quad k\to\infty,\\
  R_{\pm}^{2} &= (r+C_{\pm})^{2} + z^{2},\quad
  \psi_{\pm} = R_{\pm} + \gamma\cos^{-1}C_{\pm},\nonumber
\end{align}
and:
\begin{align}
  \label{equ:sp:I}
  I_{m} &\sim (Q_{m+2}(k,r,z) - Q_{m}(k,r,z))/4,
\end{align}
where $Q_{-m}$ and $Q_{-m-2}$ are neglected since they have no
stationary phase points and decay much faster than $Q_{m}$ and
$Q_{m+2}$ with increasing $k$.

\begin{figure*}
  \centering
  \begin{tabular}{cc}
    \includegraphics{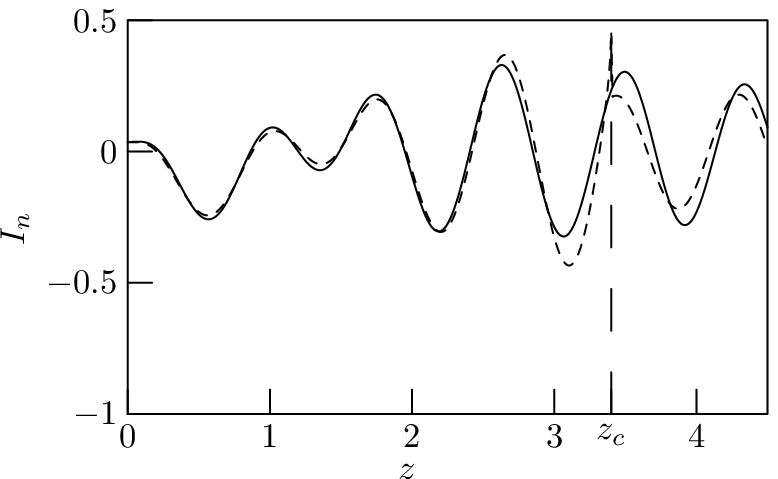} &
    \includegraphics{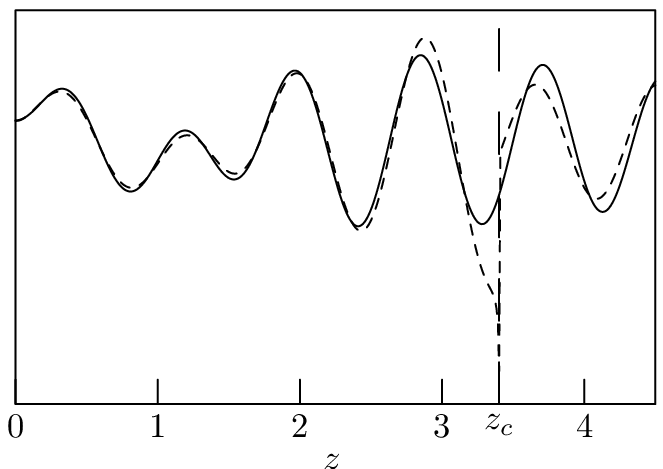}
  \end{tabular}
  \caption{Numerical and asymptotic evaluation of $I_{n}$, $r=1.125$,
    $k=8$, $n=1$: left hand side real, right hand side imaginary;
    solid line numerical evaluation, dashed line asymptotic,
    Equation~\ref{equ:sp:I}}
  \label{fig:field}
\end{figure*}

Figure~\ref{fig:field} compares the asymptotic approximation for
$I_{1}(k,r,z)$ to a numerical evaluation of the integral. The cut-off
value of $z$ for $\gamma=3/k$ is indicated and the real and imaginary
parts of the integrals are plotted separately. As $z\to z_{c}$, the
stationary phase approximation to $Q_{m+2}$ breaks down and there is a
resulting loss of accuracy. Away from this point, however, the
approximation to $I_{m}$ is accurate on both sides of $z_{c}$.

The asymptotic analysis shows that the cut-on line source modes, those
with $m<k$, radiate efficiently into a region bounded by $\pm
z_{c}(r,\gamma)$. This gives a second limit on the information
available in the acoustic field.

\subsection{Far field approximation}
\label{sec:far:field}

For completeness, we give a far field approximation of the line source
radiation integral, valid outside the region covered by the asymptotic
expansions of \S\ref{sec:sphase}. Using the standard approximation,
$R\approx R_{0} + (r-r_{2})\sin\phi$, $1/R\approx 1/R_{0}$ with
$R_{0}^{2}=r^{2}+z^{2}$:
\begin{align}
  I_{m} &\approx \J^{m}\pi\frac{\E^{\J k
      R_{0}}}{R_{0}}\frac{m+1}{k\sin\phi}
  \biggl[
      \left(r+\J\frac{(m+2)}{k\sin\phi}\right)
      J_{m+1}(k\sin\phi)\nonumber\\
      &-\J J_{m}(k\sin\phi)
      \biggr]
  \label{equ:far:field}
\end{align}
so that in the far field, $I_{m}$ decays as $k^{-1}$. 

\section{Information in spinning sound fields}
\label{sec:information}

Summarizing the results of the previous section, the nature of a
spinning sound field is seen to be determined by its wavenumber $k$
and the relation of $k$ to the set of modes contained in the
equivalent line source. The first result, that line modes with $m>k$
generate exponentially small fields, means that the acoustic field
contains a limited amount of information about the source. Such a
result has been derived previously by showing that the far field
pressure is a band-limited Fourier transform of the line source
strength\cite{carley09}, but this new result establishes an exact
limit on the information in the field, without needing a far-field
approximation.

The second result, from the stationary phase analysis, shows that the
modes which do radiate, those with $m<k$, are more efficient in some
parts of the field than in others. The higher order radiating modes
are detectable only near the source plane, with a lower radiation
efficiency at larger $z$. The only mode which radiates efficiently
over the whole field is the `plane' mode $m=0$. 

The following sections discuss some implications of these findings for
different problems.

\subsection{Low speed rotors}
\label{sec:low:speed}

One result which is of immediate interest is that, in some sense, low
speed rotors have the same acoustic field. Given that for a system of
radius $a$ rotating at angular velocity $\Omega$ the non-dimensional
wavenumber of the $n$th harmonic of the radiated field $k=n\Omega
a/c=n\mt$, with $\mt$ the source tip Mach number, low speed rotors
will have $k<1$ over the first few harmonics of the signal. This means
that the sound field is dominated by the zero order line mode and any
set of rotors of a given blade number operating at the same speed,
whatever their blade geometry, will have the same acoustic field, to
within a scaling factor. Indeed, experiments in active noise control
of noise from cooling fans\cite{gerard-berry-masson-gervais07} have
found good results by discretizing the inverse model of the fan into
three sections, i.e.\ using three degrees of freedom, for a value of
$ka\approx0.8$. The same group, in an earlier study of the
conditioning of the inversion problem, using a hemispherical
arrangement of microphones\cite{gerard-berry-masson05a} found that the
condition number reduced as frequency was increased, a finding they
ascribe to ``insufficient spatial resolution of the source for
frequency below 200\hertz.'' In the light of the analysis above, an
alternative viewpoint might be that as frequency is increased, more
line source modes are cut on and the information available in the
acoustic field becomes more nearly sufficient for source
reconstruction. 

\subsection{Source identification}
\label{sec:identification}

The original motivation for this work was the problem of identifying a
rotating source. The results of \S\ref{sec:analysis} and
\S\ref{sec:field} can be used to help show why this is a hard problem
and to indicate how it might best be approached.

The first obvious consequence of the result of \S\ref{sec:cut:off} is
that the acoustic field has a limited number of degrees of
freedom. For a field of given wavenumber $k$, no more than $k$ modes
can be detected in the field, i.e. the field has $M$ degrees of
freedom, with $M$ the largest integer $M<k$. Attempting to identify
sources using more than $M$ degrees of freedom is inherently
ill-conditioned because the modes with $m>k$ generate exponentially
small fields.

\begin{figure}
  \centering
  \includegraphics{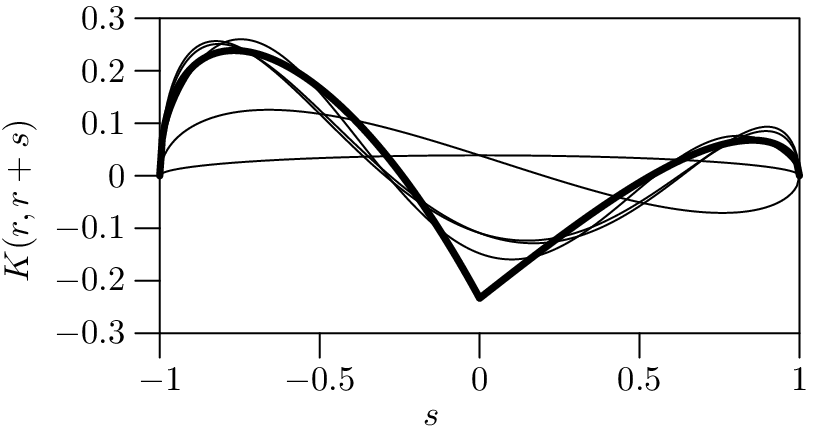}
  \caption{Line source reconstructed with different modes for $r=5/4$:
    thick line full source term $K(r,r+s)$; thin lines $K(r,r+s)$
    reconstructed using 1--5 modes.}
  \label{fig:source}
\end{figure}

Secondly, the asymptotic analysis shows that much of the information
in the field is not detected by microphones. For a given microphone
polar angle, only a subset of the $M$ radiating modes will be easily
detected. If the microphone is at polar angle $\phi$, modes with
$\gamma>\sin\phi$ may not be detected or, alternatively, the
microphone only detects radiation from modes of order $m<k/\sin\phi$. 

The effect this has on the source identification problem is shown in
Figure~\ref{fig:source}. This shows, as a thick line, $K(r,r+s)$ for
$r=5/4$, $n=2$ and $s(r_{1})\equiv1$, calculated using previously
published closed-form expressions\cite{carley99}. The thin lines show
$K$ reconstructed using the first $M$ terms of the sum in
Equation~\ref{equ:K:poly}, with $M=0,1,2,3,4$. The convergence towards
the exact value of $K$ is obvious, but it is also obvious that this
convergence is so slow that a large number of terms will be needed in
order to accurately reconstruct $K$. Given that the number of modes
which can be detected depends on wavenumber $k$, it is clear that
except at very high frequencies, only low resolution source
reconstruction is possible.

\subsection{Source resolution}
\label{sec:resolution}

The observations of the previous subsection give a pessimistic outlook
for source identification: the information necessary for accurate
reconstruction of a source is simply not available, even in theory. On
the other hand, if the objective is to reproduce an acoustic field,
for noise control, say, only limited knowledge of the source is
necessary. Looking again at Figure~\ref{fig:source}, although there is
a large difference between the exact line source and the source
produced by summation of lower order modes, the acoustic field
generated by the lower order modes with $m<k$ will be
indistinguishable from that generated by the exact source, since the
modes with $m>k$ do not contribute to the radiated sound.

\begin{figure*}
  \centering
  \begin{tabular}{cc}
    \includegraphics{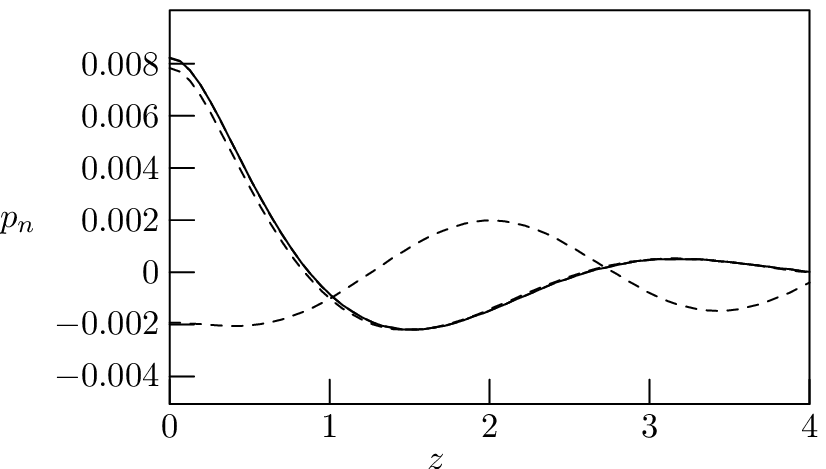} &
    \includegraphics{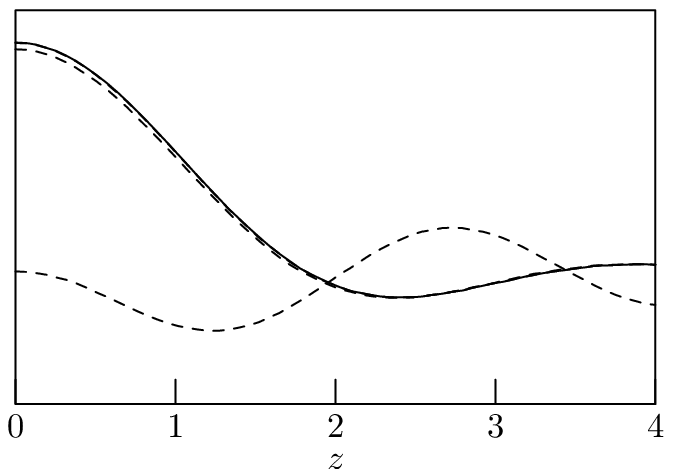}
  \end{tabular}
  \caption{Acoustic integral with exact line source (solid line) and
    lower order modes (dashed lines) for $m\leq0,2,4$, $k=2.5$,
    $r=1.25$; real part on left, imaginary part on right.}
  \label{fig:buildup}
\end{figure*}

Figure~\ref{fig:buildup} shows the acoustic field on a sideline
$r=1.25$ computed using the exact line source of
Figure~\ref{fig:source} and by summation of the field due to the lower
order modes, with $m\leq0,2,4$ shown. The wavenumber $k=2.5$, so that
cutoff begins at $m=3$. Indeed, the sum over the first three modes is
very close to the exact result, and when the first five modes are
included, the result is indistinguishable from the exact field. 

\begin{figure}
  \centering
  \includegraphics{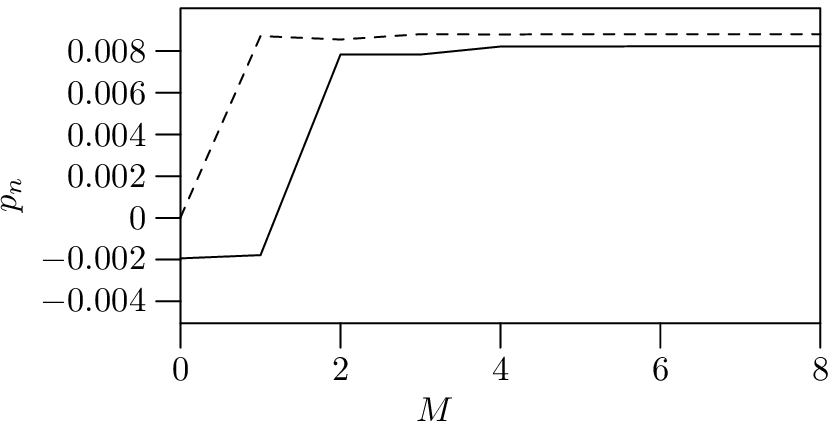}
  \caption{Acoustic integral at $z=0$ against maximum mode order $M$
    included: real part solid, imaginary part dashed.}
  \label{fig:buildup:inplane}
\end{figure}

Figure~\ref{fig:buildup:inplane} shows the development of $p_{n}$ at
$z=0$ as more modes are included in the summation. It is clear that
the sum has all but converged after the $m=2$ mode is added and is
practically unchanged after the $m=4$ mode is added, confirming the
conclusion that the higher order modes are cut-off and do not
contribute to the acoustic field, in spite of their quite high
amplitudes. The clear conclusion is that the acoustic field depends on
the lower order modes and sources which differ only in the higher
order terms of their line source decomposition cannot be distinguished
by a source identification procedure. The corollary of this statement
is that for an accurate prediction of the field, sources need only be
specified to a resolution sufficient to correctly identify the
amplitudes of the lower order modes. 

\subsection{Jet noise}
\label{sec:jets}

In a recent paper\cite{jordan-schlegel-stalnov-noack-tinney07}, Jordan
\textit{et al.}\ examine noise production by a turbulent jet using
proper orthogonal decomposition (POD) to perform a modal decomposition
of the flow and a technique called MOD (`most observable
decomposition') to decompose the acoustic far field. They find that
while~350 flow modes are needed to account for~50\% of the turbulent
kinetic energy in the flow, only~24 modes are needed to account
for~90\% of the acoustic energy. If the jet is viewed as a
distribution of disk sources along the jet axis, the results of this
paper show that we should expect that only a small fraction of the
modes will radiate noise and that in a complex source such as a jet,
the bulk of the modes will have $m>k$ and will generate exponentially
small fields. In a study of noise sources in a jet,
Freund\cite{freund01} filters the source terms to leave ``a set of
modes capable of radiating to the far field'', based on a wavenumber
criterion, but he notes that ``additional cancellation may occur due
to the radial structure of the source''. The analysis of the previous
sections offers an approach for the identification of the radial terms
which will give such cancellation.

\section{Conclusions}
\label{sec:conclusions}

An analysis of the information content of a spinning sound field has
been presented. It has been shown, on the basis of an exact analysis,
that the acoustic field around a spinning source has at most $M$
degrees of freedom with $M<k$, the acoustic wavenumber. This result
arises from the replacement of the disk source with an exactly
equivalent line source given by a sum of modes. Most of these modes
generate exponentially small acoustic fields, i.e.\ are cut off, and
the remaining, lower order, modes radiate efficiently only into
sectors of the acoustic field which become smaller as the modal order
increases. The results explain a number of features which have been
observed in the literature, including: the possibility of using low
order source models for noise control; the ill-conditioning of source
identification methods; and the low radiation efficiency of subsonic
jets. 

\bibliography{abbrev,identification,propnoise,maths,misc,solid,duct,jets}

\end{document}